# Empirical corrections and pair interaction energies in the fragment molecular orbital method

Dmitri G. Fedorov,[1,*] Jimmy C. Kromann,[2] Jan H. Jensen[2]

[1] Research Center for Computational Design of Advanced Functional Materials (CD-FMat), National Institute of Advanced Industrial Science and Technology (AIST), Central 2, Umezono 1-1-1, Tsukuba, 305-8568, Japan.

[2] Department of Chemistry, University of Copenhagen, Universitetsparken 5, 2100 Copenhagen, Denmark

Corresponding author: E-mail address: d.g.fedorov@aist.go.jp (D.G. Fedorov).

**Abstract**

The energy and analytic gradient are developed for FMO combined with the Hartree-Fock method augmented with three empirical corrections (HF-3c). The auxiliary basis set approach to FMO is extended to perform pair interaction energy decomposition analysis. The FMO accuracy is evaluated for several typical systems including 3 proteins. Pair interaction energies computed with different approaches in FMO are compared for a water cluster and protein-ligand complexes.

## 1. Introduction

A variety of low scaling methods [1] are available for calculations of large molecular systems. Among them, fragment-based methods [2,3,4,5,6,7,8,9,10,11,12] offer the advantage of not only reducing the computational cost but also providing properties of fragments, and, in many cases, interactions between them.

The fragment molecular orbital method (FMO) [13,14,15,16,17] is based on a many-body expansion [18,19], incorporating pair interaction energies (PIEs) between fragments. They provide chemical insight on the recognition of guest molecules bound to a host, for example, a ligand bound to a protein, in the form of residue fragment contributions. PIEs have been used in many applications of FMO [20].

Different PIE analyses have been suggested for FMO: a decomposition of PIEs into polarization and charge transfer [21], counterpoise basis set superposition error (BSSE) [22,23] corrections [24], decomposition of the correlated component of PIEs into localized orbital contributions [25], many-body corrections to PIEs [26], statistical correction [27], fluctuation analysis [28] and a PIE decomposition analysis (PIEDA) [29,30,31].

In this work, parametrized BSSE corrections are interfaced with FMO. PIEs obtained with this method are compared to the values from other approaches for a water cluster and protein-ligand complexes. The basis set dependence of PIEs is systematically studied for the water cluster.

## 2. Methodology

The energy expression for FMO truncated at trimer terms (FMO3) for a system divided into $N$ fragments can be written as [19]

$$E^{\mathrm{FMO3}} = \sum_{I}^{N} E'_I + \sum_{I>J}^{N} \Delta E_{IJ} + \sum_{I>J>K}^{N} \Delta E_{IJK} \tag{1}$$

where $E'_I$, $\Delta E_{IJ}$ and $\Delta E_{IJK}$ are the internal energies of fragments (monomers), dimer and trimer corrections, respectively. In FMO2, the last sum in Eq. 1 is omitted.

Fragments and their conglomerates are calculated in FMO in the presence of an embedding electrostatic potential (ESP). This ESP for fragment $X$ is calculated using the density or atomic charges of all fragments excluding $X$. Because ESP depends on the electronic state of all fragments, the fragment calculations in the presence of ESP are repeated iteratively until the embedding potential converges. After that, fragment pairs and, optionally, trimers are calculated in the presence of the ESP, which is fixed at this stage. The total properties are calculated using the many-body expansion in Eq. 1. The gradient in FMO is complicated by the dependence of ESP on the electronic state of each fragment, and one has to solve self-consistent Z-vector (SCZV) equations [32] to obtain orbital responses.

As a cheap alternative to accurate but time consuming ab initio calculations, one can employ parametrized corrections in ab initio methods. In this work, Hartree-Fock (HF) with three corrections (3c) is combined with FMO. In HF-3c [33], there is a dispersion correction D3(BJ) [34] and two BSSE corrections: short-ranged basis set (SRB) [33] and geometrical counterpoise (GCP) [35]. These corrections, independent of the electronic state, are added to the FMO-HF energy evaluated with a minimum basis set MINIX.

In PIEDA [28], the pair interactions $\Delta E_{IJ}$ in Eq. 1 are decomposed in most QM methods into 5 components: electrostatic (ES), exchange-repulsion (EX), charge transfer and mix terms (CT+mix), dispersion and remainder correlation (DI+RC) and solvent screening (SOLV).

$$\Delta E_{IJ} = \Delta E_{IJ}^{\mathrm{ES}} + \Delta E_{IJ}^{\mathrm{EX}} + \Delta E_{IJ}^{\mathrm{CT+mix}} + \Delta E_{IJ}^{\mathrm{DI+RC}} + \Delta E_{IJ}^{\mathrm{SOLV}} \tag{2}$$

For DFT, $\Delta E_{IJ}^{\mathrm{DI+RC}}$ can be decomposed as $\Delta E_{IJ}^{\mathrm{DI}} + \Delta E_{IJ}^{\mathrm{RC}}$ and for HF, $\Delta E_{IJ}^{\mathrm{DI+RC}} = \Delta E_{IJ}^{\mathrm{DI}}$. For density-functional tight-binding (DFTB) [36,28] PIEDA has different components:

$$\Delta E_{IJ} = \Delta E_{IJ}^{\mathrm{ES}} + \Delta E_{IJ}^{0} + \Delta E_{IJ}^{\mathrm{CT\cdot ES}} + \Delta E_{IJ}^{\mathrm{DI}} + \Delta E_{IJ}^{\mathrm{SOLV}} \tag{3}$$

where $\Delta E_{IJ}^{0}$ is the zeroth order Hamiltonian contribution (mainly, a non-polar term) and $\Delta E_{IJ}^{\mathrm{CT\cdot ES}}$ is the coupling of ES and CT. PIEDA for HF-3c is formulated in this work as

$$\Delta E_{IJ} = \Delta E_{IJ}^{\mathrm{ES}} + \Delta E_{IJ}^{\mathrm{EX}} + \Delta E_{IJ}^{\mathrm{CT+mix}} + \Delta E_{IJ}^{\mathrm{DI}} + \Delta E_{IJ}^{\mathrm{BS}} + \Delta E_{IJ}^{\mathrm{SOLV}} \tag{4}$$

where

$$\Delta E_{IJ}^{\mathrm{BS}} = \Delta E_{IJ}^{\mathrm{GCP}} + \Delta E_{IJ}^{\mathrm{SRB}} \tag{5}$$

The GCP and SRB terms tend to be relatively large and of the opposite sign, so it appears plausible to look at the total basis set (BS) correction as their sum. $\Delta E_{IJ}^{\mathrm{GCP}}$ and $\Delta E_{IJ}^{\mathrm{SRB}}$ are computed as the GCP and SRB energies of dimers minus monomer values, respectively.

The auxiliary basis (AB) set approach was developed [37] to improve the accuracy of FMO for large basis sets with diffuse functions. In it, the polarization effects evaluated using a smaller basis set with embedding are added to the FMO calculation with a larger basis set without embedding. FMO/AB was used to optimize a small protein [37] and to do molecular dynamics simulations in explicit solvent [38].

PIEDA/AB is formulated in this work as

$$\Delta E_{IJ} = \Delta E_{IJ}^{\mathrm{ES}} + \Delta E_{IJ}^{\mathrm{EX}} + \Delta E_{IJ}^{\mathrm{CT+mix}} + \Delta E_{IJ}^{\mathrm{DI+RC}} + \Delta E_{IJ}^{\mathrm{SOLV}} + \Delta E_{IJ}^{\mathrm{BS}} \tag{6}$$

where all PIE components except for $\Delta E_{IJ}^{\mathrm{BS}}$ are computed for the smaller basis set and the basis set correction $\Delta E_{IJ}^{\mathrm{BS}}$ is evaluated without ESP as the difference in PIEs for the larger and smaller basis sets. $\Delta E_{IJ}^{\mathrm{BS}}$ is the total basis set correction; it is possible to decompose $\Delta E_{IJ}^{\mathrm{BS}}$ into the components in Eq. 2.

It is important to understand the difference between interaction and binding energies. Interaction energies $\Delta E_{IJ}$ in FMO are pairwise binding energies between electrostatically interacting polarized fragments, whereas the binding energy $\Delta E$ for a cluster is the amount of energy gained by forming the complex from isolated non-interacting fragments. The difference between the two was described as the polarization, deformation and desolvation [39] of monomer terms. For FMO2, the binding energy can be computed as

$$\Delta E^{\mathrm{bind}} = E - \sum_{I=1}^{N} E_I^0 = \sum_{I}^{N} \Delta E_I + \sum_{I>J}^{N} \Delta E_{IJ} \tag{7}$$

where the monomer contributions $\Delta E_I = E_I' - E_I^0$ are obtained from the energies $E_I^0$ of isolated fragments. It is misleading to focus only on $\Delta E_{IJ}$ when discussing the binding, although the values of $\Delta E_{IJ}$ form a large contribution to it.

## 3. Computational details

HF-3c was interfaced to the FMO code in GAMESS [40] and parallelized using the generalized distributed data interface [41].

The following molecular systems were calculated: water cluster $(H_2O)_8$, polyalanine α-helices (α-(ALA)$_n$, $n$=10, 20 and 40), chignolin (1UAO), Trp-cage (1L2Y), crambin (1CRN) and complexes of Trp-cage with the neutral or deprotonated (anionic) *p*-phenolic acid. In the water cluster the point charge representation of ESP was used [42], and in the other systems the default ESP (two-electron integrals and point-charges for near and far fragments, respectively) was used.

Polypeptides and the water cluster were divided into 1 residue and molecule per fragment, respectively, using the hybrid orbital projection boundary treatment. C-PCM (conductor polarizable continuum model) was used at the FMO/PCM<1> level [43]. DFTB was used at the DFTB3 level [44] with 3ob parameters [45].

## 3. Results and discussion

The results for the accuracy of FMO-HF3c energy and gradient are given in Table 1 and Table 2, respectively. It can be seen that FMO2 errors are fairly large whereas FMO3 has the chemical accuracy (the errors are less than 0.5 kcal/mol). The FMO-HF3c gradient accuracy is reasonable both with respect to the numeric FMO-HF3c gradient, and to the HF-3c gradient without fragmentation.

The pair interaction energies are their components, averaged for the four strongest hydrogen bonds in $(H_2O)_8$ (previously studied with density functional theory (DFT) and DFTB in [28]) are shown in Table 3. The electrostatic term $\Delta\bar{E}^{\mathrm{ES}}$ for DFTB, MINIX and aug-cc-pVTZ is -5.6, -10.5 and -13.7 kcal/mol, respectively, i.e., the MINIX basis set has a larger electrostatic interaction than the minimum basis set in DFTB (the electrostatics in DFTB is modified via the damping and a use of distance $R$ functions more complex than $1/R$ [44]). The exchange-repulsion $\Delta\bar{E}^{\mathrm{EX}}$ and charge transfer $\Delta\bar{E}^{\mathrm{CT+mix}}$ terms for MINIX compared to aug-cc-pVTZ are smaller and larger, respectively. The solvent screening $\Delta\bar{E}^{\mathrm{SOLV}}$, determined by the solvent charges induced by the solute electrostatic potential, is very similar in all methods because the difference between methods only slightly affects the solute potential that induces the solvent charges and determines the screening.

In general, BSSE tends to predict an overbinding. MINIX is a compact minimum basis set, which means the overlap of the core orbitals of one fragment with the valence orbitals of another, being the source of BSSE, is smaller than for some medium basis sets, so that the value of the correction $\Delta\bar{E}^{\mathrm{BS}}$=1.9 kcal/mol seems reasonable.

The timings in Table 3 show that the parametrized methods DFTB and HF-3c are many orders of magnitude faster than ab initio approaches. For this relatively small system, the two parametrized methods have a similar timing, but in general DFTB is faster as all integrals are pretabulated, whereas HF-3c needs one and two-electron integrals. On the other hand, HF-3c uses no parameters for the QM (HF) part of it, and thus may be more general than DFTB.

To elucidate the efficiency of AB in describing PIEs for large basis sets, a comparative study was conducted for several combinations of basis sets. The results are shown in Table 4 and

Figure 1a. Several trends can be discerned. The cc-pV*n*Z and aug-cc-pV*n*Z families converge to each other as *n* increases, as expected. The performance of FMO/AB in reproducing full values of $\Delta\bar{E}$ is good (the largest deviation is 0.4 kcal/mol) when cc-pV*n*Z and aug-cc-pV*n*Z are mixed for the same *n* value.

For water clusters one can do full calculations with large basis sets such as aug-cc-pVQZ without the AB approach. However, for covalently connected fragments only the AB approach can be reliably used, due to SCF divergence and accuracy issues [42]. Thus, in the calculations of protein-ligand complexes below only AB calculations are reported for the basis set combination of cc-pVDZ and aug-cc-pVDZ.

The binding energies for $(H_2O)_8$ are evaluated as $\Delta E^{\text{bind}} = E - 8E^0$ from the energy of the cluster $E$ and one water molecule $E^0$ (Figure 1b). The binding energies for the aug-cc-pV*n*Z family rapidly converge, and even the *n*=2 basis has a high accuracy. The convergence pattern for the binding (Fig. 1b) and interaction energies (Fig. 1a) indicates that both monomer (polarization etc) and dimer (interaction) contributions feature a relatively slow convergence, signifying that both contributions should be taken into account for a reliable analysis of binding.

Pair interaction energies between the Trp-cage protein and the neutral and anionic ligands (previously studied in [28]) are given in Table 5. There is a good agreement between all methods, but some differences also exist. The largest deviations are for Tyr-3, Pro-18, Pro-19 and Ser-20 in the complex with the anionic ligand.

For drug discovery applications [46], the correlation is important. The PIEs for MP2 and HF-3c are shown in Fig. 2. It can be seen that the correlation is very high, with the correlation coefficients $R^2$ for MP2 of 0.999 and 0.983 for the neutral and anionic ligands, respectively. The slope for MP2 is 0.95 and 0.90 for the neutral and anionic ligand, respectively, meaning that the PIE values are systematically underestimated in HF-3c, similar to DFTB. In this system, the anionic ligand is bound into a deeper binding pocket than the neutral ligand, increasing the QM effects due to multiple short-range interactions, which may explain a somewhat worse agreement between HF-3c and MP2 for this ligand.

However, MP2/AB may show some overbinding relative to higher levels of calculation. To probe this, spin-component scaled MP2 (SCS-MP2) calculations [47] were performed which are said to improve MP2. The agreement of HF-3c to SCS-MP2 is better than to MP2, especially for the slope (increased by 0.02 for both ligands) and absolute values of PIEs. Finally, the timings of HF-3c and DFTB for the Trp-cage ligand complex on a dual 12-core 2.2 GHz Xeon node (12 cores total) were 5.9 and 0.5 min, respectively.

## 3. Conclusions

The fragment molecular orbital method up to the third order FMO3 has been extended to treat basis set superposition error corrections. The accuracy has been shown to be reasonable for a set of typical systems, in terms of the energy and analytic gradient.

The interaction energies between water molecules in a cubic cluster have been analyzed using different levels of QM calculations. All methods give a similar value of the averaged hydrogen bonding energy (from -5.9 to -6.5 kcal/mol), but there is a big discrepancy between individual components, such as the electrostatics is underestimated with the minimum basis sets in DFTB and HF-3c. Interestingly, the values of dispersion are rather different in these methods (ranging from -0.3 to -2.6 kcal/mol), despite the fact that 4 of them (HF, HF-3c, DFTB and DFT) use the same D3(BJ) model, which is parametrized depending on the method. It may be argued that either some part of dispersion is treated in the QM methods and only the "missing" part is augmented with an empirical correction, or, else, one can take the dispersion as a general empirical correction.

It has been shown that the auxiliary basis approach to FMO accurately describes the interaction energies when one combines cc-pV*n*Z and aug-cc-pV*n*Z for the same *n*. This makes it possible to deliver accurate protein-ligand interaction energies, whereas without the AB approach the calculations with diffuse functions and large basis sets in FMO for covalently connected fragments are not possible. The convergence of the interaction and binding energies has been studied as a function of the basis set size.

FMO/AB has been applied at the MP2 and HF-3c level to study protein-ligand binding for two complexes of Trp-cage with ligands. It has been shown that HF-3c interaction energies highly correlate to MP2. It is expected that HF-3c may be useful in future applications of FMO.

**Acknowledgments**

We thank the developers of the HF-3c program [48] that we integrated into GAMESS. D.G.F was supported by the JSPS KAKENHI Grant 16K05677.

a)

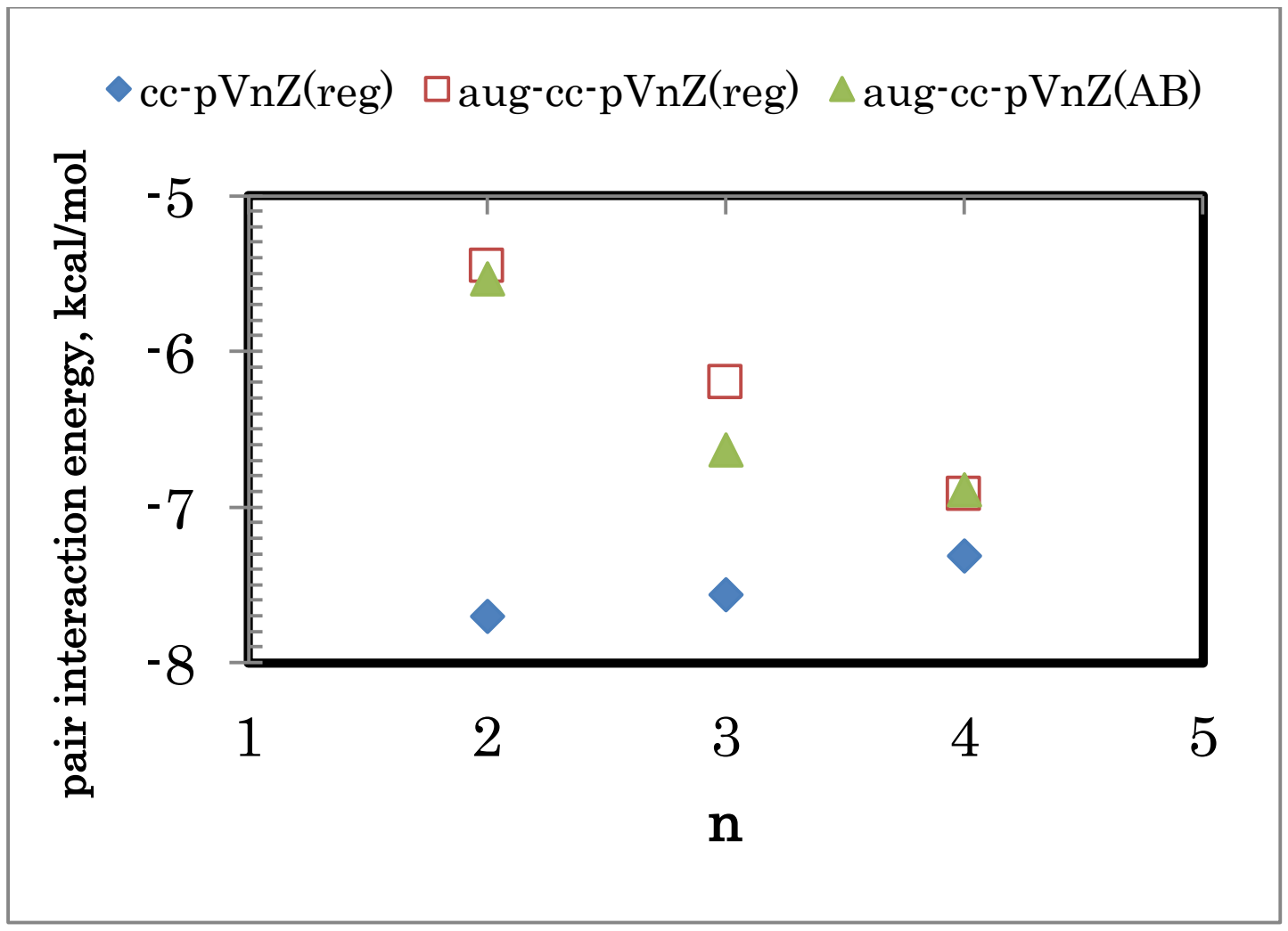

b)

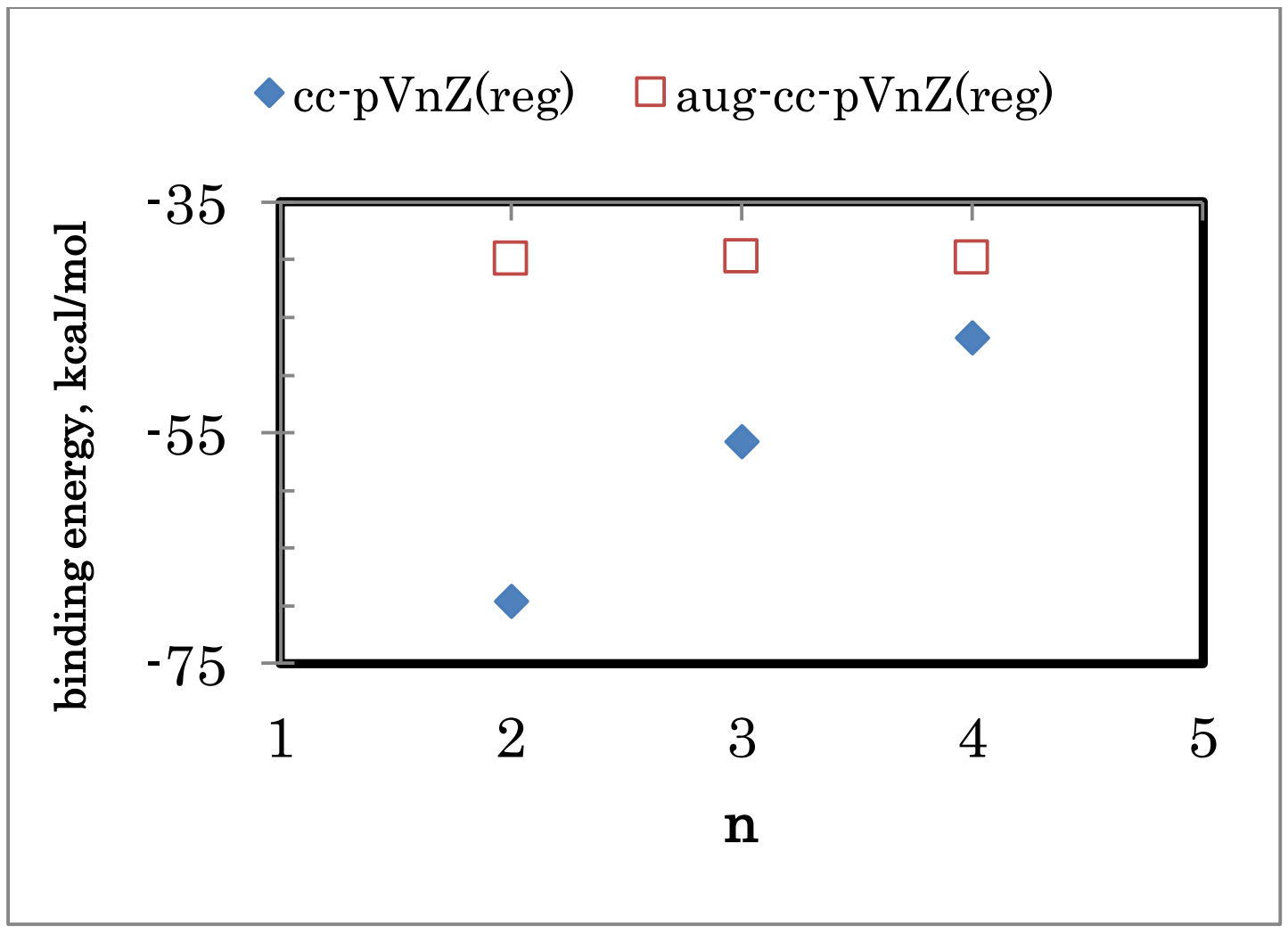

Figure 1. a) Pair interaction energies averaged over 4 strongest hydrogen bonds in $(H_2O)_8$ and b) total binding energies. “Reg” and AB refer to regular and auxiliary basis FMO calculations, respectively. In the latter, both cc-PV*n*Z and aug-cc-pV*n*Z basis sets are used.

**a) neutral ligand**

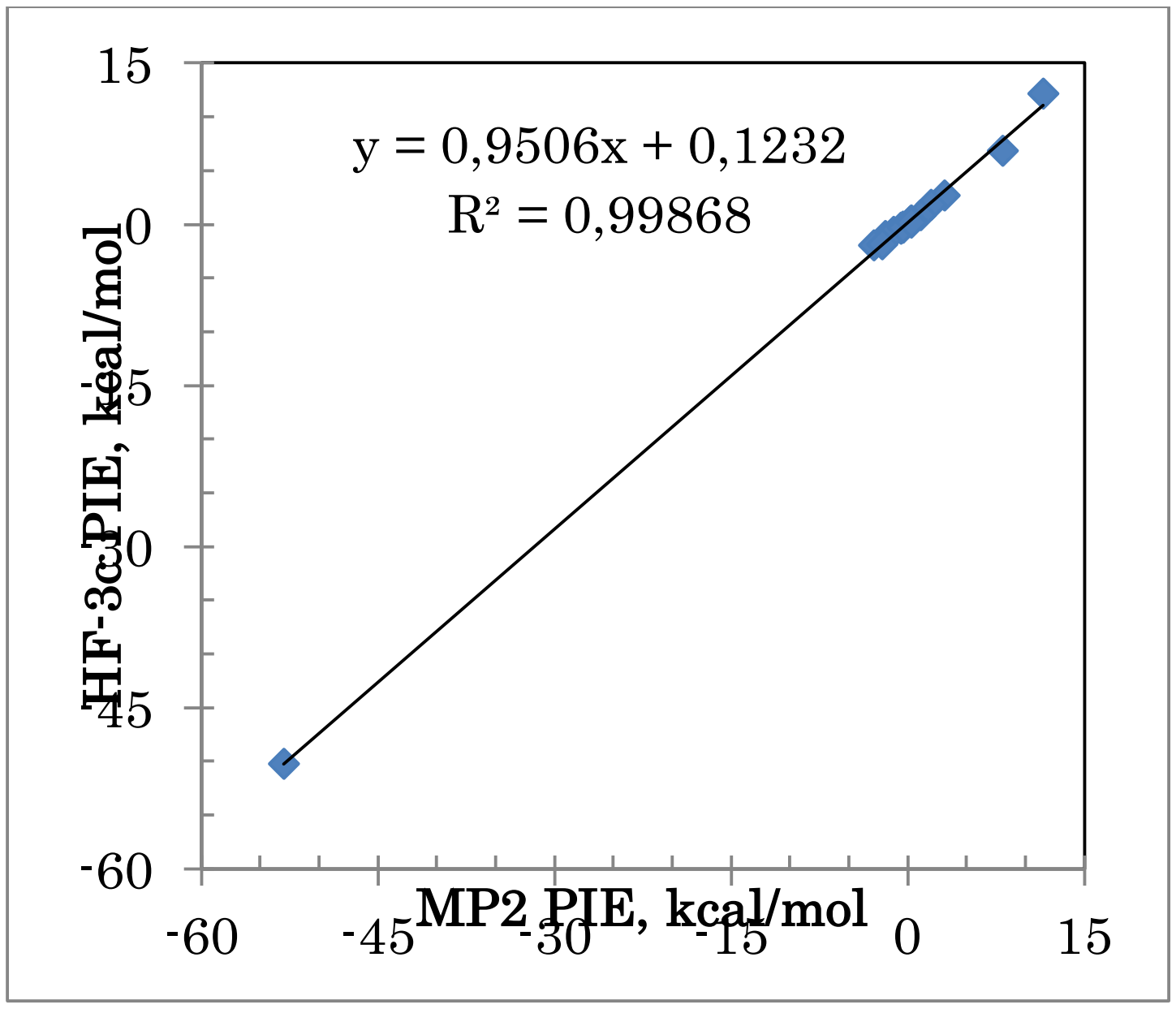

**b) anionic ligand**

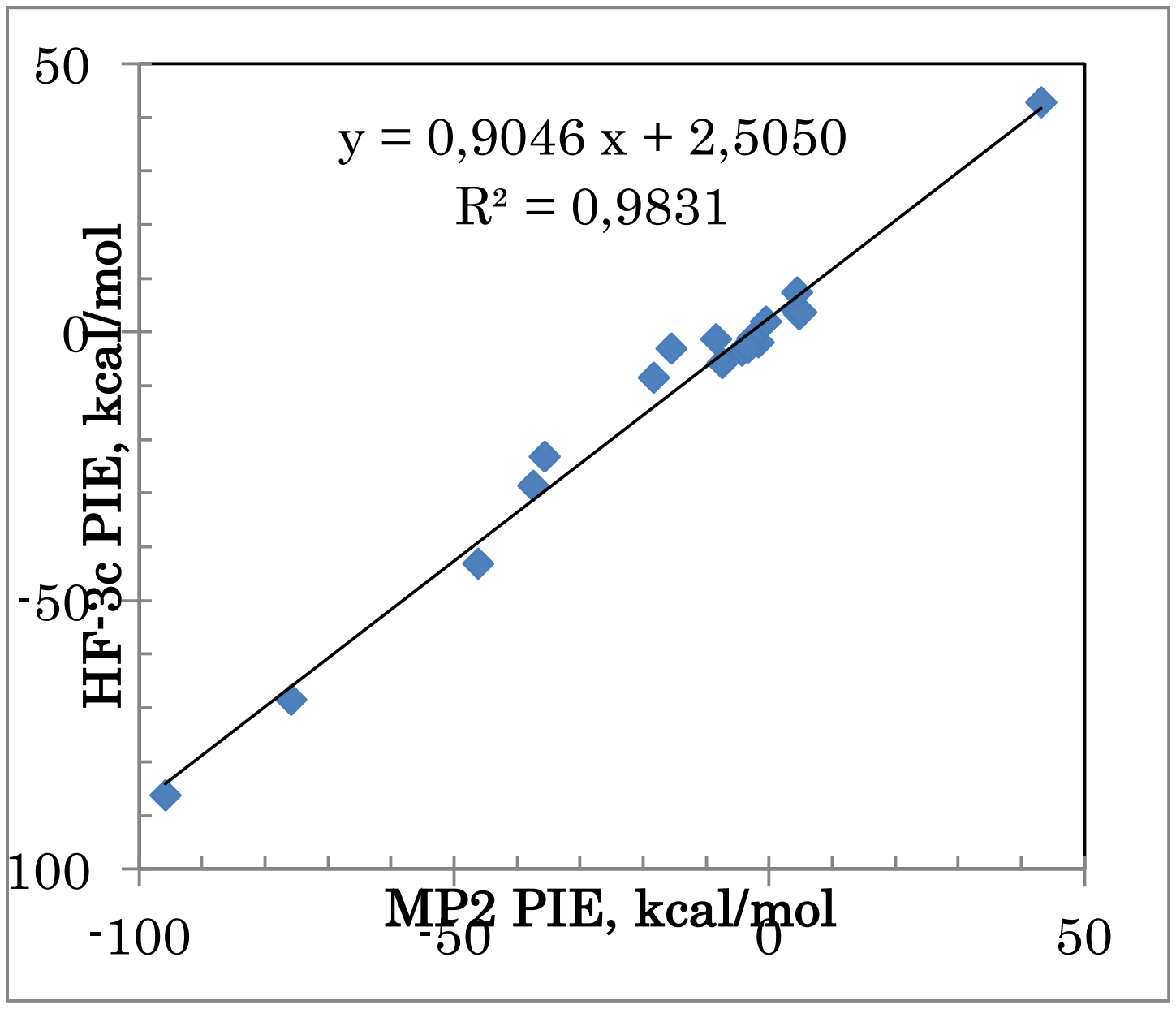

Figure 2. Correlation of the MP2/AB(cc-PV*n*Z:aug-cc-pV*n*Z) and HF-3c residue-ligand pair interaction energies (PIEs) for the complexes of Trp-cage with the a) neutral and b) anionic ligands.

**Tables**

**Table 1. Accuracy of the FMO-HF3c energies (kcal/mol) vs full HF-3c.[a]**

| System | $N^{at}$ | $N$ | $N^*$ | FMO2 | FMO3 |
|---|---|---|---|---|---|
| $\alpha$-(ALA)$_{10}$ | 112 | 10 | 0 | 1.25 | -0.02 |
| $\alpha$-(ALA)$_{10}$ | 212 | 20 | 0 | 2.80 | 0.02 |
| $\alpha$-(ALA)$_{10}$ | 412 | 40 | 0 | 5.83 | 0.09 |
| Chignolin (1UAO) | 138 | 10 | 4 | 1.37 | -0.05 |
| Trp-cage (1L2Y) | 304 | 20 | 5 | 8.18 | 0.13 |
| Crambin (1CRN) | 642 | 39 | 6 | 20.02 | 0.44 |

For systems composed of $N^{at}$ atoms divided into $N$ fragments, of which $N^*$ are charged.

**Table 2. Accuracy (hartree/bohr) of the analytic FMO2-HF3c gradient in terms of maximum (MAX) element deviations and RMSD.**

| | versus numeric FMO gradient | | versus analytic unfragmented gradient | |
|---|---|---|---|---|
| System | MAX | RMSD | MAX | RMSD |
| $\alpha$-(ALA)$_{10}$ | 0.00011 | 0.00003 | 0.00052 | 0.00014 |
| Chignolin (1UAO) | 0.00016 | 0.00004 | 0.00103 | 0.00017 |

**Table 3. Pair interactions $\Delta\bar{E}$ and their components (kcal/mol) in FMO2, averaged for the 4 strongest hydrogen bonds in $(H_2O)_8$.[a]**

| method | $T$ | $\Delta\bar{E}^{ES}$ | $\Delta\bar{E}^{0}$ | $\Delta\bar{E}^{CT\cdot ES}$ | | $\Delta\bar{E}^{DI}$ | $\Delta\bar{E}^{SOLV}$ | $\Delta\bar{E}$ |
|---|---|---|---|---|---|---|---|---|
| DFTB/D[b] | 0.01 | -5.6 | -0.6 | -0.2 | | -0.5 | 0.9 | -6.1 |
| | | | | | | | | |
| | | $\Delta\bar{E}^{ES}$ | $\Delta\bar{E}^{EX}$ | $\Delta\bar{E}^{CT+MIX}$ | | $\Delta\bar{E}^{DI}$ | $\Delta\bar{E}^{SOLV}$ | $\Delta\bar{E}$ |
| HF/D[b] | 2.2 | -13.7 | 11.8 | -2.7 | | -2.2 | 0.9 | -5.9 |
| | | | | | | | | |
| | | $\Delta\bar{E}^{ES}$ | $\Delta\bar{E}^{EX}$ | $\Delta\bar{E}^{CT+MIX}$ | $\Delta\bar{E}^{RC}$ | $\Delta\bar{E}^{DI}$ | $\Delta\bar{E}^{SOLV}$ | $\Delta\bar{E}$ |
| CAM-B3LYP/D[b] | 4.8 | -13.2 | 13.9 | -4.1 | -3.0 | -0.3 | 0.8 | -6.0 |
| | | | | | | | | |
| | | $\Delta\bar{E}^{ES}$ | $\Delta\bar{E}^{EX}$ | $\Delta\bar{E}^{CT+MIX}$ | | $\Delta\bar{E}^{RC+DI}$ | $\Delta\bar{E}^{SOLV}$ | $\Delta\bar{E}$ |
| MP2[b] | 3.5 | -13.7 | 11.8 | -2.7 | | -2.5 | 0.9 | -6.2 |
| CCSD(T) [b] | 137.4 | -13.7 | 11.8 | -2.7 | | -2.6 | 0.9 | -6.3 |
| | | | | | | | | |
| | | $\Delta\bar{E}^{ES}$ | $\Delta\bar{E}^{EX}$ | $\Delta\bar{E}^{CT+MIX}$ | $\Delta\bar{E}^{BS}$ | $\Delta\bar{E}^{DI}$ | $\Delta\bar{E}^{SOLV}$ | $\Delta\bar{E}$ |
| HF-3c | 0.01 | -10.5 | 9.1 | -5.9 | 1.9 | -1.7 | 1.0 | -6.0 |

[a] DFTB3 with 3ob parameters; HF-3c with MINIX, other methods with aug-cc-pVTZ. HF, HF-3c and DFT(B) methods use D3(BJ) as the dispersion (D) correction. The timings $T$ (min) are for a dual 12-core 2.2 GHz Xeon node (12 cores total).

[b] Results are taken from [28] but averaged in this work for the 4 strongest rather than for all 12 hydrogen bonds in [28].

**Table 4. Pair interactions $\Delta\overline{E}$ and their components (kcal/mol) in FMO2-MP2/AB, averaged for the 4 strongest hydrogen bonds in $(H_2O)_8$.[a]**

| BS1 | BS2 | $\Delta\overline{E}^{ES}$ | $\Delta\overline{E}^{EX}$ | $\Delta\overline{E}^{CT+MIX}$ | $\Delta\overline{E}^{BS}$ | $\Delta\overline{E}^{RC+DI}$ | $\Delta\overline{E}^{SOLV}$ | $\Delta\overline{E}$ |
|---|---|---|---|---|---|---|---|---|
| cc-pVDZ | cc-pVDZ | -11.5 | 9.6 | -4.3 | 0 | -2.4 | 0.9 | -7.7 |
| cc-pVTZ | cc-pVTZ | -13.4 | 10.5 | -3.0 | 0 | -2.7 | 1.0 | -7.6 |
| cc-pVQZ | cc-pVQZ | -14.3 | 11.3 | -2.8 | 0 | -2.6 | 1.0 | -7.3 |
| | | | | | | | | |
| cc-pVDZ | aug-cc-pVDZ | -11.5 | 9.6 | -4.3 | 2.2 | -2.4 | 0.9 | -5.5 |
| aug-cc-pVDZ | aug-cc-pVDZ | -12.8 | 11.7 | -3.1 | 0 | -2.2 | 0.8 | -5.5 |
| | | | | | | | | |
| cc-pVDZ | aug-cc-pVTZ | -11.5 | 9.6 | -4.3 | 2.2 | -2.4 | 0.9 | -5.5 |
| cc-pVTZ | aug-cc-pVTZ | -13.4 | 10.5 | -3.0 | 0.9 | -2.7 | 1.0 | -6.6 |
| aug-cc-pVTZ | aug-cc-pVTZ | -13.7 | 11.8 | -2.7 | 0 | -2.5 | 0.9 | -6.2 |
| | | | | | | | | |
| cc-pVTZ | aug-cc-pVQZ | -13.4 | 10.5 | -3.0 | 1.0 | -2.7 | 1.0 | -6.6 |
| cc-pVQZ | aug-cc-pVQZ | -14.3 | 11.3 | -2.8 | 0.4 | -2.6 | 1.0 | -6.9 |
| aug-cc-pVQZ | aug-cc-pVQZ | -14.8 | 12.0 | -2.6 | 0 | -2.5 | 0.9 | -6.9 |

[a] Auxiliary basis (AB) calculations are the same as regular FMO if the two basis sets BS1 and BS2 are identical.

**Table 5. Pair interaction energies (kcal/mol) in the complexes of Trp-cage.[a]**

| | | Neutral ligand | | | | Anionic ligand | | | |
|---|---|---|---|---|---|---|---|---|---|
| residue | *Q* | MP2 | SCS-MP2 | HF-3c | DFTB/D [a] | MP2 | SCS-MP2 | HF-3c | DFTB/D [a] |
| Asn-1 | 1 | 3.1 | 3.1 | 2.7 | 2.6 | -75.8 | -74.5 | -68.5 | -70.2 |
| Leu-2 | 0 | -0.6 | -0.6 | -0.3 | -0.4 | -35.7 | -33.5 | -23.1 | -33.2 |
| Tyr-3 | 0 | -0.6 | -0.6 | -0.4 | -0.6 | 4.5 | 4.7 | 3.9 | 12.4 |
| Ile-4 | 0 | -0.5 | -0.5 | -0.3 | -0.5 | 4.7 | 4.7 | 3.3 | 6.3 |
| Gln-5 | 0 | -1.9 | -1.9 | -1.1 | -1.6 | -37.4 | -36.1 | -28.5 | -33.0 |
| Trp-6 | 0 | -2.9 | -2.9 | -2.0 | -2.1 | -8.4 | -7.3 | -1.2 | -8.4 |
| Leu-7 | 0 | -1.2 | -1.2 | -0.7 | -1.3 | -7.4 | -7.4 | -5.8 | -0.5 |
| Lys-8 | 1 | 11.5 | 12.1 | 12.1 | 12.5 | -46.2 | -46.2 | -43.1 | -45.4 |
| Asp-9 | -1 | -53.0 | -51.5 | -50.3 | -41.9 | 43.2 | 43.3 | 42.9 | 42.4 |
| Gly-10 | 0 | 2.0 | 2.0 | 1.8 | 0.1 | -4.3 | -4.3 | -3.4 | -1.5 |
| Gly-11 | 0 | 1.1 | 1.1 | 0.8 | 0.7 | -2.5 | -2.5 | -1.9 | -3.1 |
| Pro-12 | 0 | -0.4 | -0.4 | -0.2 | -0.7 | 5.0 | 5.0 | 3.8 | 5.1 |
| Ser-13 | 0 | 0.2 | 0.2 | 0.2 | 0.1 | -1.6 | -1.6 | -1.8 | -1.0 |
| Ser-14 | 0 | 0.3 | 0.3 | 0.1 | 0.6 | -2.6 | -2.6 | -1.0 | -3.4 |
| Gly-15 | 0 | 0.3 | 0.3 | 0.3 | 0.0 | -3.3 | -3.3 | -2.8 | -2.0 |
| Arg-16 | 1 | 8.1 | 8.1 | 6.8 | 7.2 | -95.8 | -92.8 | -86.2 | -83.6 |
| Pro-17 | 0 | -0.4 | -0.4 | -0.3 | -0.5 | -0.5 | -0.3 | 2.0 | 3.1 |
| Pro-18 | 0 | -0.4 | -0.4 | -0.3 | -0.2 | -18.4 | -16.9 | -8.5 | -8.9 |
| Pro-19 | 0 | 0.3 | 0.3 | 0.1 | 0.1 | -15.6 | -12.8 | -3.1 | -12.1 |
| Ser-20 | -1 | -2.2 | -2.2 | -1.9 | -2.1 | 4.5 | 4.6 | 7.5 | 11.9 |

[a] DFTB3-D3(BJ)/3ob results are taken from [28]. Using AB (cc-pVDZ:aug-cc-pVDZ) for MP2. *Q* is the residue charge.

## References

[1] A. V. Akimov, O. V. Prezhdo, Chem. Rev. 115 (2015) 5797.

[2] M. S. Gordon, S. R. Pruitt, D. G. Fedorov, L. V. Slipchenko, Chem. Rev. 112 (2012) 632.

[3] P. Otto, J. Ladik, Chem. Phys. 8 (1975) 192.

[4] J. Gao, J. Phys. Chem. B 101 (1997) 657.

[5] T. Fang, Y. Li, S. Li, WIREs: Comput. Mol. Sc. 7 (2017) e1297.

[6] J. Liu, J. Z. H. Zhang, X. He, Phys. Chem. Chem. Phys. 18 (2016) 1864.

[7] R. Kobayashi, R. Amos, M. A. Collins, J. Phys. Chem. A 121 (2017) 334.

[8] H. Yu, H. R. Leverentz, P. Bai, J. I. Siepmann, D. G. Truhlar, J. Phys. Chem. Lett. 5 (2014) 660.

[9] N. Sahu, S. R. Gadre, J. Chem. Phys. 144 (2016) 114113.

[10] K. V. J. Jose, K. Raghavachari, J. Chem. Theory Comput. 11 (2015) 950.

[11] J. Liu, J. M. Herbert, J. Chem. Theory Comput. 12 (2016) 572.

[12] P. K. Gurunathan, A. Acharya, D. Ghosh, D. Kosenkov, I. Kaliman, Y. Shao, A. I. Krylov, L. V. Slipchenko, J. Phys. Chem. B 120 (2016) 6562.

[13] K. Kitaura, E. Ikeo, T. Asada, T. Nakano, M. Uebayasi, Chem. Phys. Lett. 313 (1999) 701.

[14] D. G. Fedorov, K. Kitaura, J. Phys. Chem. A 111 (2007) 6904.

[15] D. G. Fedorov, T. Nagata, K. Kitaura, Phys. Chem. Chem. Phys. 14 (2012) 7562.

[16] S. Tanaka, Y. Mochizuki, Y. Komeiji, Y. Okiyama, K. Fukuzawa, Phys. Chem. Chem. Phys. 16 (2014) 10310.

[17] D. G. Fedorov, WIREs: Comput. Mol. Sc. 7 (2017) e1322.

[18] D. G. Fedorov, K. Kitaura, J. Chem. Phys. 120 (2004) 6832.

[19] D. G. Fedorov, N. Asada, I. Nakanishi, K. Kitaura, Acc. Chem. Res. 47 (2014) 2846.

[20] I. Morao, D. G. Fedorov, R. Robinson, M. Southey, A. Townsend-Nicholson, M. J. Bodkin, A. Heifetz, J. Comput. Chem. 38 (2017) 1987.

[21] Y. Mochizuki, K. Fukuzawa, A. Kato, S. Tanaka, K. Kitaura, T. Nakano, Chem. Phys. Lett. 410 (2005) 247.

[22] D. W. Schwenke, D. G. Truhlar, J. Chem. Phys. 82 (1985) 2418.

[23] J. C. Faver, Z. Zheng, K. M. Merz, Jr. J. Chem. Phys. 135 (2011) 144110.

[24] T. Ishikawa, T. Ishikura, K. Kuwata, J. Comput. Chem. 30 (2009) 2594.

[25] T. Ishikawa, Y. Mochizuki, S. Amari, T. Nakano, H. Tokiwa, S. Tanaka, K. Tanaka, Theor. Chem. Acc. 118 (2007) 937.

[26] C. Watanabe, K. Fukuzawa, Y. Okiyama, T. Tsukamoto, A. Kato, S. Tanaka, Y. Mochizuki, T. Nakano, J. Mol. Graph. Modell. 41 (2013) 31.

[27] S. Tanaka, C. Watanabe, Y. Okiyama, Chem. Phys. Lett. 556 (2013) 272.

[28] D. G. Fedorov, K. Kitaura, J. Phys. Chem. A 122 (2018) 1781.

[29] D. G. Fedorov, K. Kitaura, J. Comput. Chem. 28 (2007) 222.

[30] D. G. Fedorov, K. Kitaura, J. Phys. Chem. A 116 (2012) 704.

[31] M. C. Green, D. G. Fedorov, K. Kitaura, J. S. Francisco, L. V. Slipchenko, J. Chem. Phys. 138 (2013) 074111.

[32] T. Nagata, K. Brorsen, D. G. Fedorov, K. Kitaura, M. S. Gordon, J. Chem. Phys. 134 (2011) 124115.

[33] R. Sure, S. Grimme, J. Comput. Chem. 34 (2013) 1672.

[34] S. Grimme, S. Ehrlich, L. Goerigk, J. Comput. Chem. 32 (2011) 1456.

[35] H. Kruse, S. Grimme, J. Chem. Phys. 134 (2012) 154101.

[36] Y. Nishimoto, D. G. Fedorov, S. Irle, J. Chem. Theory Comput. 10 (2014) 4801.

[37] D. G. Fedorov, K. Kitaura, Chem. Phys. Lett. 597 (2014) 99.

[38] S. R. Pruitt, K. R. Brorsen, M. S. Gordon, Phys. Chem. Chem. Phys. 17 (2015) 27027.

[39] D. G. Fedorov, K. Kitaura, J. Phys. Chem. A 120 (2016) 2218.

[40] M. W. Schmidt, K. K. Baldridge, J. A. Boatz, S. T. Elbert, M. S. Gordon, J. H. Jensen, S. Koseki, N. Matsunaga, K. A. Nguyen, S. Su, et al. J. Comput. Chem. 14 (1993) 1347.

[41] D. G. Fedorov, R. M. Olson, K. Kitaura, M. S. Gordon, S. Koseki, J. Comput. Chem. 25 (2004) 872.

[42] D. G. Fedorov, L. V. Slipchenko, K. Kitaura, J. Phys. Chem. A 114 (2010) 8742.

[43] T. Nagata, D. G. Fedorov, H. Li, K. Kitaura, J. Chem. Phys. 136 (2012) 204112.

[44] M. Gaus, Q. Cui, M. Elstner, WIREs Comput. Mol. Sci. 4 (2014) 49.

[45] M. Gaus, A. Goez,. M. Elstner, J. Chem. Theory Comput. 9 (2013) 338.

[46] M. P. Mazanetz, E. Chudyk, D. G. Fedorov, Y. Alexeev, Applications of the fragment molecular orbital method to drug research. In *Computer aided drug discovery*. W. Zhang (Ed.) Springer, New York, 2016, pp. 217-255.

[47] S. Grimme J. Chem. Phys. 118 (2003) 9095.

[48] https://www.chemie.uni-bonn.de/pctc/mulliken-center/software. Accessed April 20, 2018.